%% file: eprint_arxiv.tex
\documentclass[12pt]{article}
\usepackage{graphicx}
\usepackage{amsmath}

\newcommand\pubnumber{CMS-CR-2018/262}
\newcommand\pubdate{\today}

\def\institute{Institute of Experimental Particle Physics\\
Karlsruhe Institute of Technology, 76131 Karlsruhe, GERMANY}
\def\Title#1{\begin{center} {\Large #1 } \end{center}}
\def\Author#1{\begin{center}{ \sc #1} \end{center}}
\def\Address#1{\begin{center}{ \it #1} \end{center}}

\newcommand\pubblock{\rightline{\begin{tabular}{l} \pubnumber\\
         \pubdate  \end{tabular}}}
\newenvironment{Abstract}{\begin{quotation}  }{\end{quotation}}
\newenvironment{Presented}{\begin{quotation} \begin{center} 
             PRESENTED AT\end{center}\bigskip 
      \begin{center}\begin{large}}{\end{large}\end{center} \end{quotation}}

\input econfmacros.tex
\begin{document}
\begin{titlepage}
\pubblock

\vfill
\Title{Measurement of the single top quark and antiquark production cross sections in the $t$ channel and their ratio at 13 TeV}
\vfill
\Author{Denise M\"uller\\on behalf of the CMS Collaboration}
\Address{\institute}
\vfill
\begin{Abstract}
The electroweak production in the $t$ channel is the most dominant production mode of single top quarks at the LHC. The ratio of the cross sections of the top quark and antiquark production provides an insight into the inner structure of the proton and is therefore suitable to study different parton distribution function predictions. This process can also be used for a direct measurement of the absolute value of the CKM matrix element $V_\textrm{tb}$. The most recent measurement of the CMS Collaboration is presented with the 2016 data set of the LHC Run II at a center-of-mass energy of 13\,TeV. Events with one muon or one electron are considered in this analysis. Lepton-flavor dependent multivariate discriminators are applied to separate signal from background events.
\end{Abstract}
\vfill
\begin{Presented}
$11^\textrm{th}$ International Workshop on Top Quark Physics\\
Bad Neuenahr, Germany, September 16--21, 2018
\end{Presented}
\vfill
\end{titlepage}
\def\thefootnote{\fnsymbol{footnote}}
\setcounter{footnote}{0}

\section{Introduction}

\begin{figure}[h]
\centering
\includegraphics[width=0.3\textwidth]{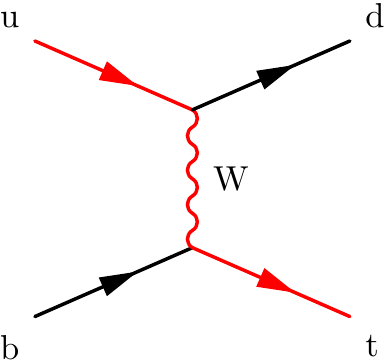}\hspace{3em}
\includegraphics[width=0.3\textwidth]{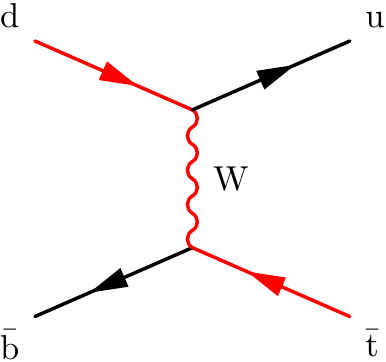}
\caption{Feymnan diagrams at Born level for the production of single top quarks (left) and single top antiquarks (right).}
\label{fig1}
\end{figure}

In the standard model (SM) of elementary particles, single top quarks are produced via the electroweak interaction. 
Out of three possible production modes, the $t$-channel process is the dominant mechanism at the LHC, corresponding to approximately 70\% of the single top quark cross section at a center-of-mass energy of 13\,TeV. 
In this production mode, the top quark is produced through the interaction of a W boson and a bottom quark, together with a light quark in forward direction. 
Figure~\ref{fig1} illustrates the production of a single top quark and single top antiquark. 
The flavor of the initial light quark defines whether a top quark or a top antiquark is produced. 
By calculating the ratio of the cross sections of these two processes, $R_{t\textrm{-ch}} = \sigma_{t\textrm{-ch,t}}/\sigma_{t\textrm{-ch,}\bar{\textrm{t}}}$, an insight into the proton structure is provided and different parton distribution functions (PDFs) can be studied.
The theoretical cross sections at 13\,TeV for top quark, top antiquark and total $t$-channel production at next-to-leading order, calculated with \textsc{hathor}~2.1~\cite{Aliev:2010zk,Kant:2014oha}, are

\begin{align*}
\sigma_{t\textrm{-ch,t}} &= 136.0^{+4.1}_{-2.9} (\mathrm{scale}) \pm 3.5 (\mathrm{PDF +} \alpha_{\mathrm{s}}) ~\mathrm{pb}, \\
\sigma_{t\textrm{-ch,}\bar{\textrm{t}}} &= 81.0^{+2.5}_{-1.7} (\mathrm{scale}) \pm 3.2 (\mathrm{PDF +} \alpha_{\mathrm{s}}) ~\mathrm{pb}, \\
\sigma_{t\textrm{-ch,t+}\bar{\textrm{t}}} &= 217.0^{+6.6}_{-4.6} (\mathrm{scale}) \pm 6.2 (\mathrm{PDF +} \alpha_{\mathrm{s}}) ~\mathrm{pb}.
\end{align*}

Using these values, the predicted cross section ratio is $R_{t\textrm{-ch}} = 1.68 \pm 0.02$, including the uncertainties due to renormalization and factorization scales, and PDF+$\alpha_{\mathrm{s}}$.

\section{Measurement}

The data set recorded in 2016 by the CMS detector~\cite{Chatrchyan:2008aa} at the LHC is analyzed, corresponding to an integrated luminosity of $35.9\,\mathrm{fb}^{-1}$. Events are selected in different categories of jet and b jet multiplicity. 
As the final state of the single top quark $t$-channel process consists of one light quark recoiling against the top quark and a bottom quark from the top quark decay, the signal category requires two jets, one of which identified as originating from a bottom quark. 
In addition, two dedicated control categories are defined to constrain the most dominant background process, top quark pair production. 
All event categories require one isolated muon or electron.
To account for the neutrino and to suppress as many QCD multijet events as possible, events with one muon (electron) are required to have a transverse W boson mass of $m_\mathrm{T}^\mathrm{W} > 50\,\mathrm{GeV}$ (missing transverse momentum of $p_\mathrm{T}^\mathrm{miss} > 30\,\mathrm{GeV}$).

\begin{figure}[h]
\centering
\includegraphics[width=0.4\textwidth]{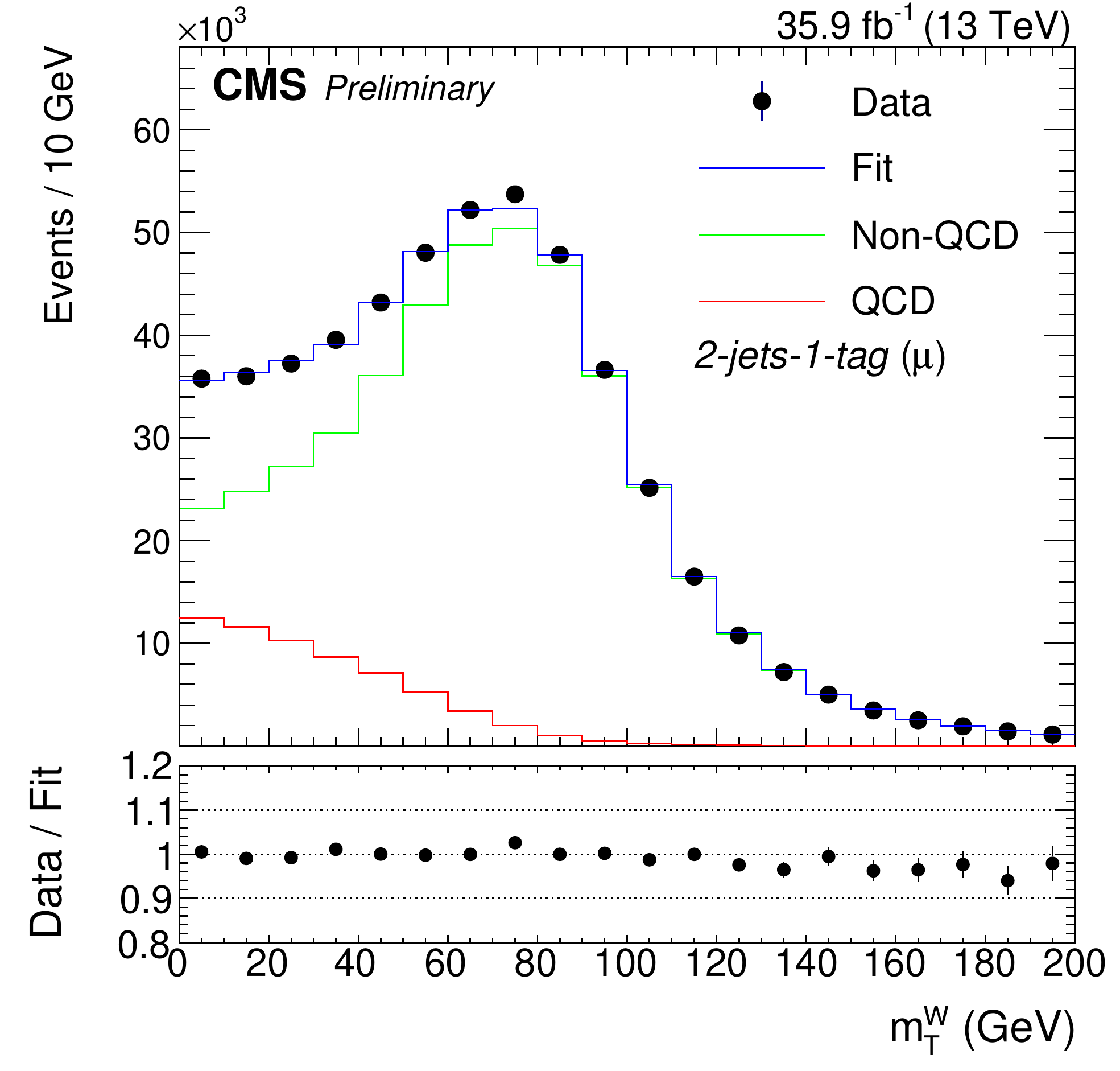}\hspace{1em}
\includegraphics[width=0.4\textwidth]{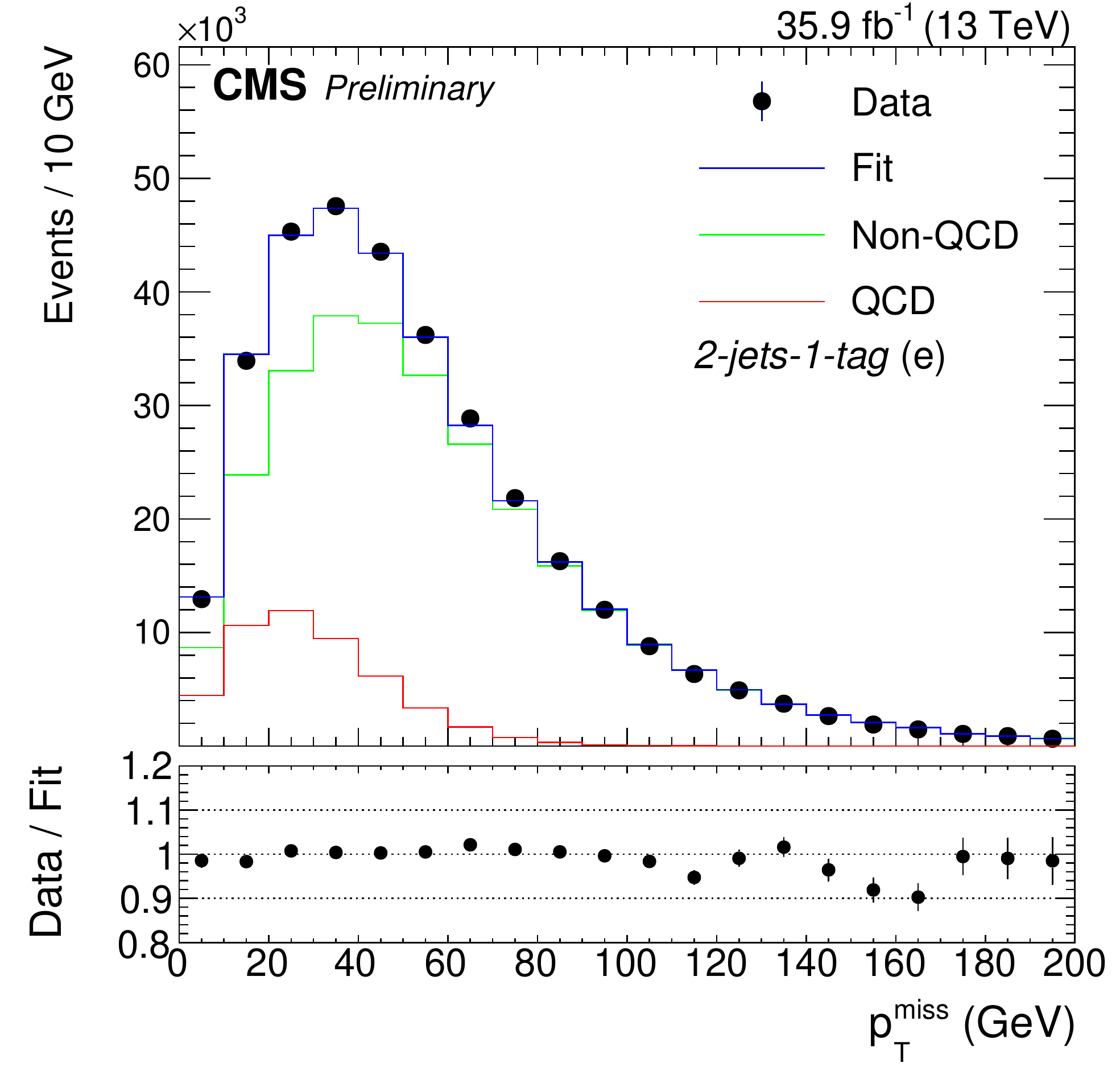}
\caption{Result of the QCD estimation for events with a muon (left) and with an electron (right). Figures taken from~\cite{CMS:2018nfg}.}
\label{fig2}
\end{figure}

The remaining QCD contribution is estimated using a data-driven approach, as the simulation of the QCD multijet simulation is not reliable, while all other processes are taken from simulation. 
A two-template fit to a discriminating variable, i.\,e., $m_\mathrm{T}^\mathrm{W}$ ($p_\mathrm{T}^\mathrm{miss}$) for muons (electrons), is performed to estimate the yield of the QCD contribution. 
The first template corresponds to the QCD template taken from data in a sideband region enriched with QCD multijet events, while the second template contains all simulated non-QCD processes including the signal. 
The result of the fit is shown in Figure~\ref{fig2}.

\begin{figure}[h]
\centering
\includegraphics[width=0.4\textwidth]{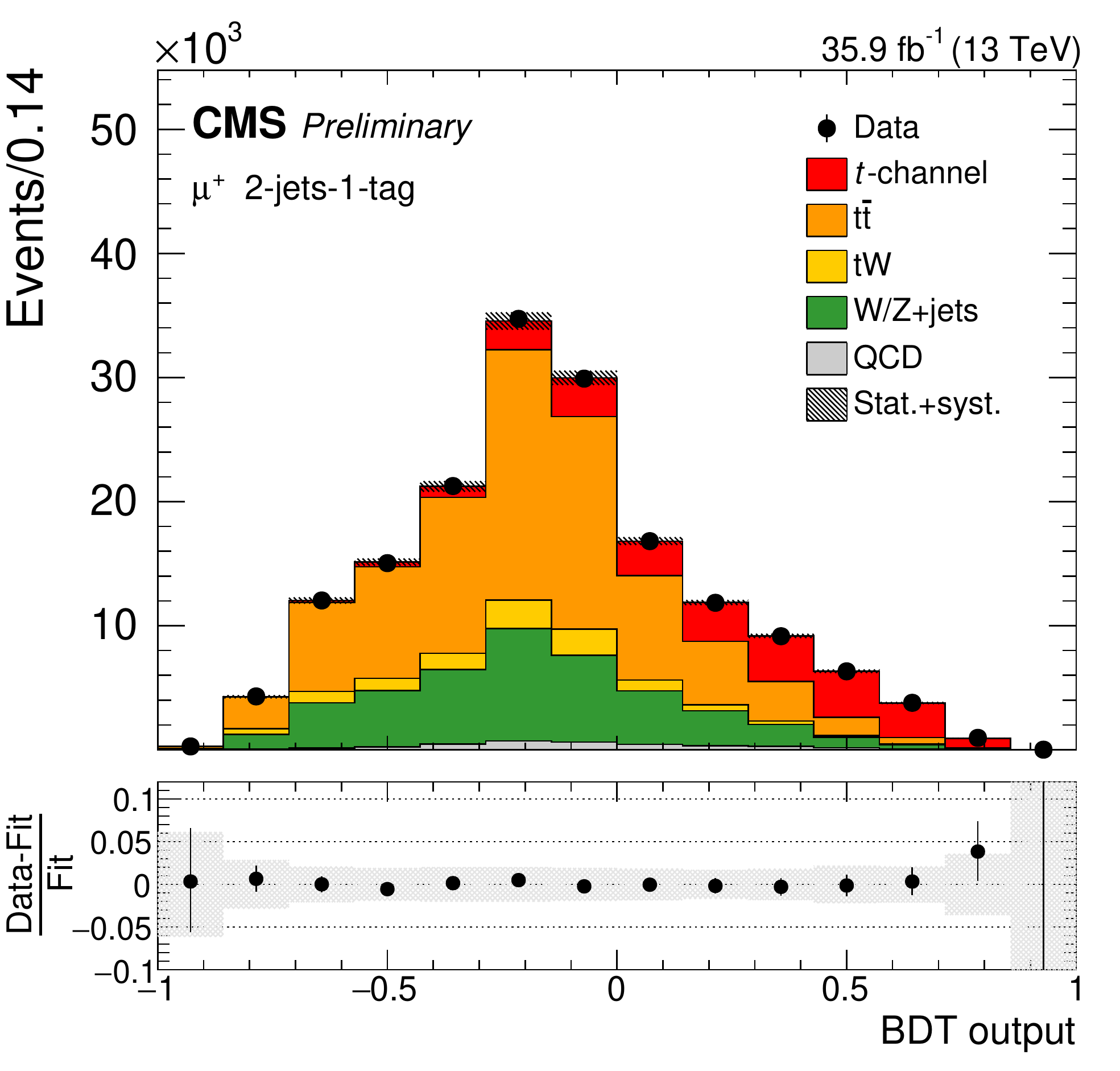}\hspace{1em}
\includegraphics[width=0.4\textwidth]{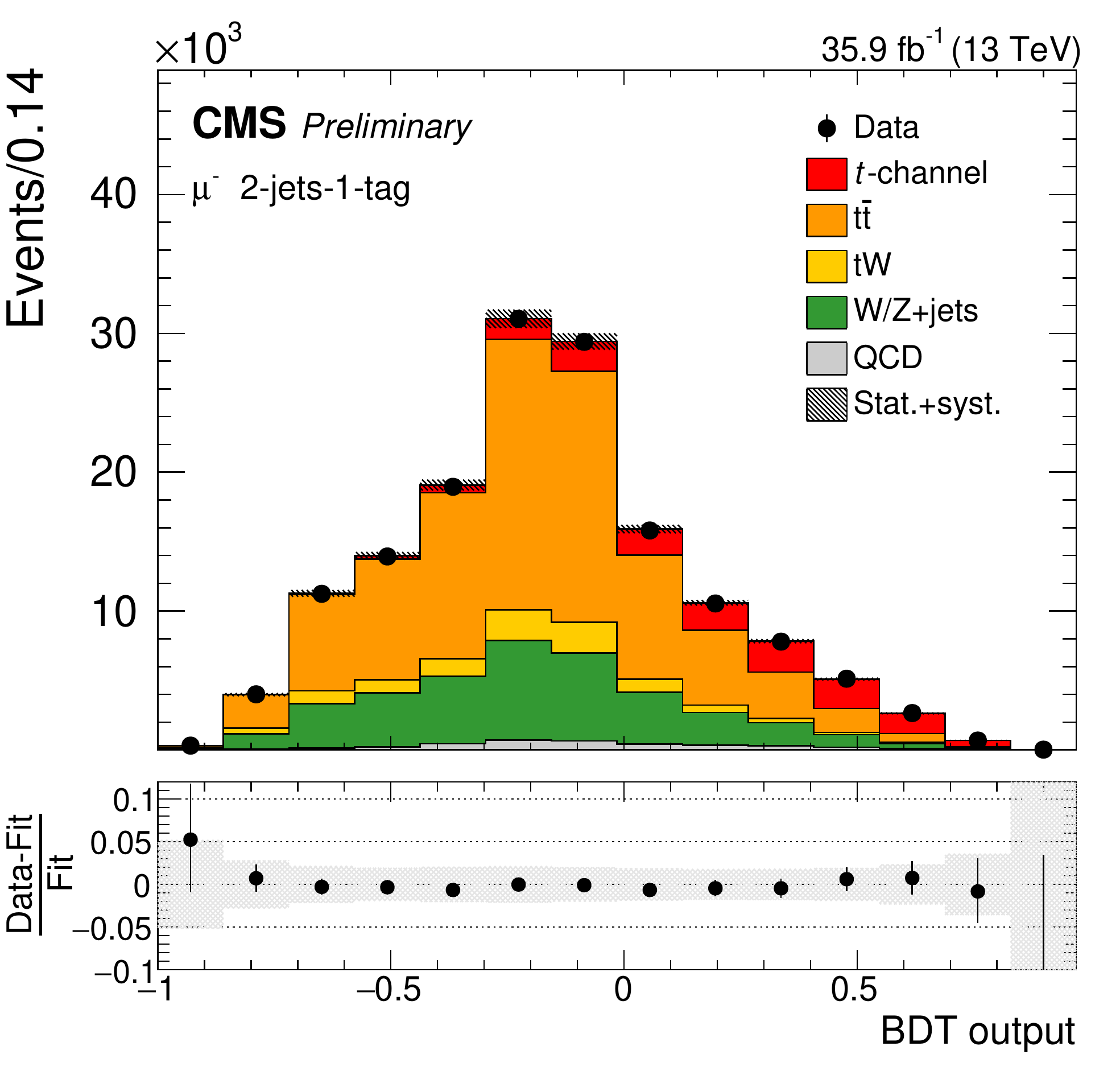}
\caption{Post-fit BDT output distribution in the signal category for positively (left) and negatively charged muons (right). Good agreement between the measured data and the simulation is found. Figures taken from~\cite{CMS:2018nfg}.}
\label{fig3}
\end{figure}

To separate the signal process from the background processes, a boosted decision tree (BDT) is employed for each lepton flavor. 
The BDTs are trained in the signal category, where the top quark pair production ($\mathrm{t\bar{t}}$), and the W+jets and QCD processes are considered as background. 
For the BDT training, different kinematic input variables, e.\,g., light-quark jet $|\eta|$, top quark mass, and invariant mass of the two jets, are used. 
The BDT output is applied to the signal and control categories for each lepton flavor and charge, and a maximum likelihood fit is performed simultaneously on these twelve BDT output distributions. 
The free parameters of this fit are the cross sections of the $t$-channel top quark and antiquark production, $\sigma_{t\textrm{-ch,t}}$ and $\sigma_{t\textrm{-ch,}\bar{\textrm{t}}}$, and their ratio $R_{t\textrm{-ch}}$.
The post-fit BDT output distributions of the signal category in the muon channel are shown in Figure~\ref{fig3}.

\section{Results}

The measured cross section ratio and the measured cross sections for top quark, top antiquark, and total $t$-channel production are

\begin{align*}
R_{t\textrm{-ch}} &= 1.65 \pm 0.02 \mathrm{(stat)} \pm 0.04 \mathrm{(syst)}, \\
\sigma_{t\textrm{-ch,t}} &= 136.3 \pm 1.1 \mathrm{(stat)} \pm 20.0 \mathrm{(syst)} ~\mathrm{pb}, \\
\sigma_{t\textrm{-ch,}\bar{\textrm{t}}} &= 82.7 \pm 1.1 \mathrm{(stat)} \pm 13.0 \mathrm{(syst)} ~\mathrm{pb}, \\
\sigma_{t\textrm{-ch,t+}\bar{\textrm{t}}} &= 219.0 \pm 1.5 \mathrm{(stat)} \pm 33.0 \mathrm{(syst)} ~\mathrm{pb}.
\end{align*}

\begin{figure}[h]
\centering
\includegraphics[width=0.65\textwidth]{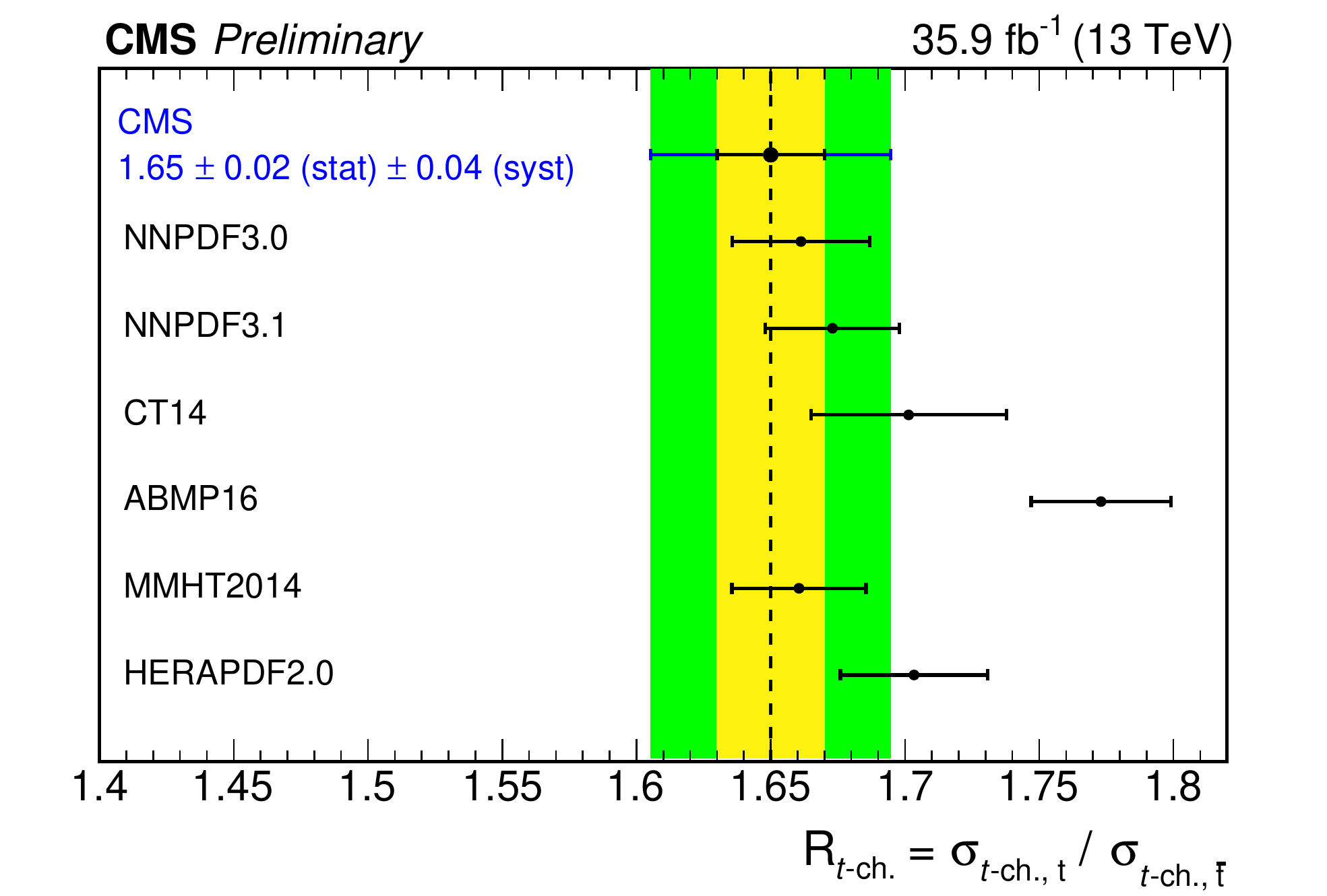}
\caption{Comparison of the measured $t$-channel cross section ratio with different PDF predictions. The uncertainties in the PDF sets include statistical uncertainty and the uncertainties due to renormalization and factorization scales and top quark mass. Agreement with most PDF sets is observed. Figure taken from~\cite{CMS:2018nfg}.}
\label{fig4}
\end{figure}

The results are dominated by systematic uncertainties: for the cross section measurements, the signal parton shower scale and the jet energy scale are the most important uncertainties, while for the ratio the signal PDF uncertainty is crucial.
The measured cross section ratio is compared with the predictions of different PDF sets, which is shown in Figure~\ref{fig4}. In addition, the CKM matrix element $|V_\mathrm{tb}|$ can be calculated with the total measured $t$-channel cross section, taking a form factor for a possible anomalous Wtb vertex into account ($f_\mathrm{LV}=1$ for SM):

\begin{equation*}
|f_\mathrm{LV} V_\mathrm{tb}| = 1.00 \pm 0.05 \mathrm{(exp)} \pm 0.02 \mathrm{(theo)}.
\end{equation*} 

All measured quantities are compatible with the SM predictions. Further details can be found in Ref.~\cite{CMS:2018nfg} and updated results in Ref.~\cite{Sirunyan:2018rlu}.

\end{document}

%% file: econfmacros.tex
\def\beq{\begin{equation}}
\def\eeq#1{\label{#1}\end{equation}}
\def\eeqn{\end{equation}}

\def\beqa{\begin{eqnarray}}
\def\eeqa#1{\label{#1}\end{eqnarray}}
\def\eeqan{\end{eqnarray}}

\let\bar=\overbar

\def\Dslash{\not{\hbox{\kern-4pt $D$}}}
\def\dslash{\not{\hbox{\kern-2pt $\del$}}}

\def\msb{{\bar{\ssstyle M \kern -1pt S}}}